\documentclass[pra,twocolumn,superscriptaddress,amssymb,amsmath,amsmath,showpacs]{revtex4}
\usepackage{color}
\usepackage{graphicx}
\usepackage{epstopdf}
\usepackage{multirow}
\usepackage{natbib}
\usepackage{wasysym}
\usepackage{hyperref} 

\hypersetup{%
   pdfpagemode=None, 
   pdfstartpage=1,
   pdfstartview=FitH,
   pdfmenubar=true,
   pdftoolbar=true,
   colorlinks = true,
   linkcolor=blue,
   citecolor=blue,
   bookmarksopen=false
 }

\newcommand{\Op}[1]{\boldsymbol{\mathsf{\hat{#1}}}}

\newcommand{\Fkt}[1]{\,\mathsf {#1}}

\def\openone{\leavevmode\hbox{\small1\kern-3.3pt\normalsize1}}

\ifx\Tr\renewcommand{\Tr}{\Fkt{Tr}} 
\else\newcommand{\Tr}{\Fkt{Tr}}
\fi

\begin{document}

\title{Optimized production of ultracold ground-state molecules: 
  Stabilization employing potentials with ion-pair character and 
  strong spin-orbit coupling}

\author{Micha\l~Tomza}
\affiliation{Faculty of Chemistry, University of Warsaw, Pasteura 1, 
  02-093 Warsaw, Poland}
\affiliation{Theoretische Physik, Universit\"at Kassel, 
  Heinrich-Plett-Str. 40, 34132 Kassel, Germany}
\author{Michael H. Goerz}
\affiliation{Theoretische Physik, Universit\"at Kassel, 
  Heinrich-Plett-Str. 40, 34132 Kassel, Germany}
\author{Monika Musia\l}
\affiliation{Institute of Chemistry, University of Silesia, 
  Szkolna 9, 40-006 Katowice, Poland}
\author{Robert Moszynski}
\affiliation{Faculty of Chemistry, University of Warsaw, 
  Pasteura 1, 02-093 Warsaw, Poland}
\author{Christiane P. Koch}
\email{christiane.koch@uni-kassel.de}
\affiliation{Theoretische Physik, Universit\"at Kassel, 
  Heinrich-Plett-Str. 40, 34132 Kassel, Germany}

\date{\today}

\begin{abstract}
We discuss the production of ultracold molecules in their electronic
ground state by photoassociation employing electronically excited
states with ion-pair character and strong spin-orbit interaction. A
short photoassociation laser pulse drives a non-resonant three-photon
transition for alkali atoms colliding in their lowest triplet
state. The excited state wave packet is transferred to the ground
electronic state by a second laser pulse, driving a resonant
two-photon transition. After analyzing the transition matrix elements
governing the stabilization step, we discuss the efficiency of
population transfer using 
transform-limited and linearly chirped laser pulses. Finally, we
employ optimal control theory to determine the most efficient stabilization
pathways. We find that the stabilization efficiency can be increased
by one and two orders of magnitude when using linearly chirped and
optimally shaped laser pulses, respectively.  
\end{abstract}


\maketitle


\section{Introduction}
\label{sec:intro}

Photoassociation, forming  molecules  from ultracold atoms using laser
light~\cite{JonesRMP06}, is a prime candidate for coherent
control which utilizes the wave nature of matter in order to steer
a process, such as formation of a chemical bond, toward a desired
target~\cite{RiceBook,ShapiroBook}.  
At very low temperature, the delicate build-up of constructive and
destructive interference between different quantum pathways is not
hampered by thermal averaging. The basic tool for coherent control are
short laser pulses that can be shaped in their amplitude, phase and
polarization. They can drive both adiabatic and non-adiabatic 
photoassociation dynamics~\cite{KochChemRev12}.

A particular feature of photoassociation at very low temperatures is
the excitation of an atom pair at fairly large interatomic
separations~\cite{JonesRMP06}. This results from a compromise between 
the atom pair density in the electronic ground state, highest
at large interatomic separations, and population of excited state
bound levels with reasonable binding energies, that increase with
decreasing interatomic separations. Therefore, the free-to-bound 
transition matrix elements are largest for photoassociation at
separations of 50$\,$a$_0$ to 150$\,$a$_0$ with corresponding
detunings of less than about $20\,$cm$^{-1}$. Although these matrix
elements are optimally chosen, they are several orders of magnitude
smaller than those for the 
excitation of atoms. This poses a problem for photoassociation with
short laser pulses which inherently have a large bandwidth. As soon as
the wings of the pulse spectrum overlap with the atomic resonance,
atoms instead of bound levels are excited~\cite{KochPRA06b}, and 
subsequent spontaneous emission depletes the trapped
sample~\cite{SalzmannPRA06,BrownPRL06}. In photoassociation
experiments using broadband femtosecond laser pulses, the pulse
spectrum therefore needed to be cut to suppress excitation of the atomic
resonance~\cite{SalzmannPRL08}. The sharp spectral cut yields long
wings of the temporal pulse profile, and the ensuing photoassociation
dynamics were dominated by transient Rabi oscillations of extremely
weakly bound molecules caused by the long tail of the
pulse~\cite{SalzmannPRL08,MullinsPRA09,MerliPRA09,McCabePRA09}. While
it was gratifying to see that femtosecond photoassociation is
feasible~\cite{SalzmannPRL08}, larger binding energies and vibrational
instead of electronic dynamics are required to produce stable
molecules in their electronic ground state~\cite{KochChemRev12}.  

Femtosecond photoassociation at very low temperature corresponds to driving
a narrow-band transition with a broad-band laser. This can be achieved
by employing multi-photon rather than one-photon
transitions~\cite{MeshulachNature98}. The high peak powers of
femtosecond laser pulses easily allow for driving multi-photon
transitions, and multi-photon control schemes have
been demonstrated for both
weak~\cite{MeshulachNature98,MeshulachPRA99} and strong
laser pulses~\cite{TralleroPRA05,WeinachtPRL06,TralleroPRA07}. In the
weak-field regime, perturbation theory shows that optical interference
of two or more photons can be used to completely suppress
absorption~\cite{MeshulachNature98,MeshulachPRA99}. For intermediate
intensities, higher order perturbation theory can be employed to
obtain rational pulse shapes that allow to control the
absorption~\cite{GandmanPRA07,GandmanPRA07b,ChuntonovPRA08,ChuntonovJPB08,ZoharPRL08}. In
the strong-field regime, 
dynamic Stark shifts drive the transition out of resonance. This can be
countered by a linear chirp of the pulse which compensates the phase
accumulated due to the Stark shift. Aditionally adjusting the amplitude of the
pulse to guarantee a $\pi$ or $2\pi$ pulse controls the
absorption~\cite{TralleroPRA05,WeinachtPRL06,TralleroPRA07}. 
These control schemes can be applied to femtosecond photoassociation
in order to suppress the excitation of atoms~\cite{KochFaraday09}. 
Multi-photon transitions can also be utile for femtosecond
photoassociation at high
temperature~\cite{RybakPRL11,RybakFaraday11}. There the main advantage
derives from the larger flexibility in transition energies, obtained
when combining two or more photons, and the new selection
rules. The disadvantage of high temperatures is the low initial
coherence, or quantum purity, of the thermal ensemble of
atoms. However, femtosecond photoassociation can generate
rovibrational coherence by Franck-Condon
filtering~\cite{RybakPRL11}. The long-standing goal of coherently 
controlling bond formation~\cite{RonnieDancing89} 
has thus become within reach also for high temperatures. 

Besides the possibility of driving a narrow-band transition, 
multi-photon femtosecond photoassociation also allows for accessing
highly excited electronic states that may have significant ion-pair
character. Such states are expected to be well suited for the
formation of stable molecules in their electronic ground state due to
the peculiar shapes of the potential energy curves obtained when an
ion-pair state crosses covalent ones~\cite{BanCPL01,BanEPL04}. 

These
effects become most significant for heavy atoms with strong spin-orbit
interaction. The coupling of two (or more) electronic states leads to
strong mixing of the rovibrational levels provided the coupling
becomes resonant~\cite{AmiotPRL99}. The wavefunctions of such strongly
mixed levels display peaks at all the four classical turning
points. This leads to 
 large transition matrix elements for both photoassociation
and subsequent stabilization to the electronic ground
state~\cite{DionPRL01}. For homonuclear diatomics, usually several
neighbouring vibrational levels are affected by the resonant
coupling~\cite{PechkisPRA07,FiorettiJPB07}, making
them particularly suitable for short pulse photoassociation and
stabilization since a laser pulse addresses a wave packet, not just a
single level. In the case of heteronuclear
molecules, the resonantly perturbed levels are typically isolated within
the vibrational spectrum. However, the peaks at the inner turning
points are so large that stabilization into deeply bound levels of the
ground state well~\cite{GhosalNJP09,LondonoPRA09} all the way down to
$v''=0$ for SrYb~\cite{TomzaPCCP11} becomes feasible in a single
step. Strong spin-orbit interaction furthermore allows for 
singlet-triplet conversion~\cite{SagePRL05,NdongPRA10}. 

\begin{figure}[t]
  \centering
  \includegraphics[width=\columnwidth]{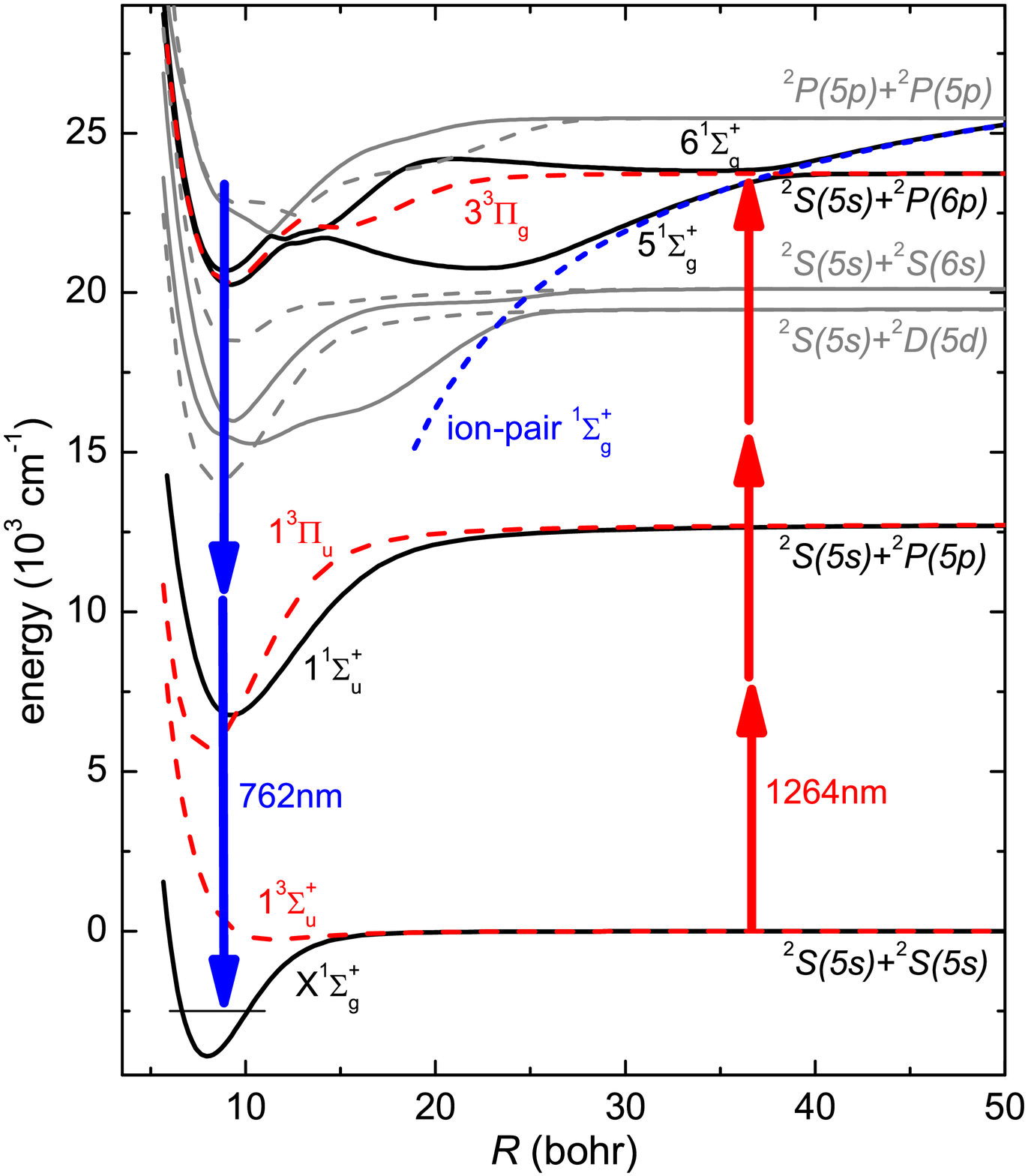}
  \caption{(Color online) Proposed scheme for multi-photon
    photoassociation and subsequent stabilization producing 
    ultracold Rb$_2$ molecules their electronic ground state.}
  \label{fig:Scheme}
\end{figure}
Here, we combine all these features in a study of 
short-pulse multi-photon photoassociation into highly excited states
with significant ion-pair character and strong spin-orbit
interaction. Our envisioned scheme for off-resonant three-photon
photoassociation and subsequent resonant two-photon stabilization is
displayed in Fig.~\ref{fig:Scheme}. The potential energy curves shown
in Fig.~\ref{fig:Scheme} as well as the spin-orbit couplings and
transition matrix elements were calculated with state of the art
\textit{ab initio} methods. The interaction of the atom pair with the
laser pulses is modelled non-perturbatively. 
A photoassociation pulse excites a
pair of rubidium atoms, colliding in their lowest triplet state, into
the manifold of electronically excited states just below the
$5s+6p_{1/2}$ threshold. The shape of the potential energy curves
close to the 
Franck-Condon radius for photoassociation is dominated by the $1/R$
behavior of the ion-pair state. Note that excitation into the ion-pair
state is dipole-forbidden. The three-photon photoassociation indicated
in Fig.~\ref{fig:Scheme} is facilitated by the covalent states that
are coupled to the ion-pair state. Upon its creation by the
photoassociation pulse, the wave packet rolls down the $1/R$ slope,
being subject to spin-orbit oscillations between different electronic
states of the manifold at the same time. Once the wave packet reaches
short interatomic separations, a stabilization pulse catches it,
transferring it to the electronic ground state via a resonant
transition into the intermediate $A^1\Sigma_u^+$-$b^3\Pi_u$
states. The strong spin-orbit coupling in both the highly excited and
intermediate state affords a conversion of the molecules to purely
singlet character in the final state. Our emphasis in the present
study is on the intricate dynamics of the photoassociated wave packet
in the electronic manifold below the $5s+6p_{1/2}$ asymptote and the
stabilization to the electronic ground state. We employ rationally
shaped laser pulses as well as optimal control theory to determine the
maximum stabilization efficiencies, taking standard constraints of
pulse shaping experiments into account. The application of optimal
control theory to the photoassociation step requires a theoretical 
description that accounts for the initial incoherent ensemble of atoms
and will be presented elsewhere. 

The plan of our paper is as follows. The theoretical framework is
presented in Sec.~\ref{sec:theory}, describing the model in
Sec.~\ref{subsec:H}, and details of the electronic structure calculations
and optimal control theory in Sec.~\ref{subsec:elstructure}
and~\ref{subsec:oct}, respectively.  The creation of the
photoassociated wave packet, representing the initial state for the
stabilization step, and its dynamics in the coupled manifold of
electronic states below the $5s+6p_{1/2}$ asymptote is studied in
Sec.~\ref{sec:ini}. The stabilization efficiency of transform-limited
and linearly chirped pulses is investigated in
Sec.~\ref{sec:anapulse}. Optimally shaped pulses driving the excited
state wave packet into the electronic ground state and their
efficiency are discussed in Sec.~\ref{sec:oct}. We conclude in
Sec.~\ref{sec:concl}.


\section{Theoretical framework}
\label{sec:theory}


\subsection{Hamiltonian}
\label{subsec:H}

\begin{figure}[t!]
\centering
\includegraphics[width=\columnwidth]{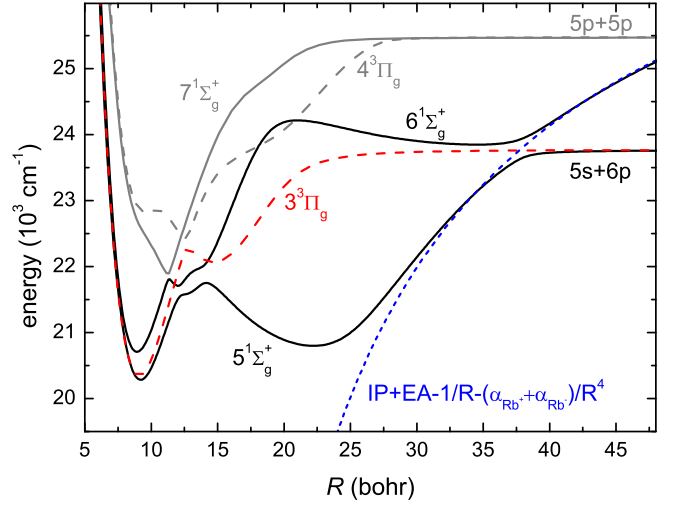}
\caption{(Color online) The $(5)^1\Sigma_g^+$, $(6)^1\Sigma_g^+$, and
  $(3)^3\Pi_g$ electronic states relevant for initial wave packet
  dynamics and ion-pair curve.} 
\label{fig:5s+6p}
\end{figure}

We consider a pair of $^{85}$Rb atoms, held at a temperature of
100$\,\mu$K, typical for magneto-optical traps, colliding in the
$a^3\Sigma_u^+$ lowest triplet state. The formation of molecules by
photoassociation and stabilization of the excited state molecules to
the eletronic ground state are treated separately. First, 
a photoassociation laser pulse drives a three-photon transition, red
detuned with respect to the ${}^2S(5s)+{}^2P_{1/2}(6p)$ 
asymptote, creating a molecular wave packet in
the manifold of the 
$(5)^1\Sigma_g^+$, $(6)^1\Sigma_g^+$, $(7)^1\Sigma_g^+$, $(3)^3\Pi_g$, and
$(4)^3\Pi_g$ electronically excited states that partially have an ion-pair
character, cf. Fig.~\ref{fig:5s+6p}. 
The states in this manifold are coupled by spin-orbit interaction and
non-adiabatic radial coupling matrix elements.
In our calcuations, only the
$(5)^1\Sigma_g^+$, $(6)^1\Sigma_g^+$, and $(3)^3\Pi_g$ components of the
photoassociated wave packet turned out to be significant~\footnote{
Specifically, the population of the $(7)^1\Sigma_g^+$ and $(4)^3\Pi_g$
states at instant of time was smaller by a factor 100 or more than the
population of the $(5)^1\Sigma_g^+$, $(6)^1\Sigma_g^+$, and
$(3)^3\Pi_g$ states.}. 
Neglecting the $(7)^1\Sigma_g^+$ and $(4)^3\Pi_g$ states, the
Hamiltonian describing the three-photon photoassociation  reads
\begin{widetext}
\begin{equation}
\label{eq:H_pump}    
\Op{H}_{pump}(t)=\begin{pmatrix}  
  \Op{H}^{a^3\Sigma_u^+}(R) & 0 & 
  \epsilon^{*}(t)^3\chi^{(3)}(\omega_L,R) & 0 \\
  0 & \Op{H}^{(5)^1\Sigma_g^+}(R) & \xi_3(R) & A(R) \\
  \epsilon(t)^3\chi^{(3)}(R) & \xi_3(R) & 
  \Op{H}^{(3)^3\Pi_g}(R) - \xi_4(R) & \xi_5(R)\\
  0 & A(R) & \xi_5(R) & \Op{H}^{(6)^1\Sigma_g^+}(R)
\end{pmatrix}
\end{equation}
\end{widetext}
in the (three-photon) rotating wave approximation. 
In Eq.~\eqref{eq:H_pump}, $\Op{H}^{^{2S+1}|\Lambda|}$ denotes the
Hamiltonian for the nuclear 
motion in the ${^{2S+1}|\Lambda|}$ electronic state, 
\begin{equation}
  \Op{H}^{^{2S+1}|\Lambda|}=\Op{T}+V^{^{2S+1}|\Lambda|}(R)+
  \omega_S^{^{2S+1}|\Lambda|}(t,R)+\Delta_{\omega_L},
\end{equation} 
with the kinetic energy operator given by $\Op{T}=\frac{\hbar^2}{2\mu
}\frac{d^2}{dR^2}$, $\mu$ the reduced mass and 
$V^{^{2S+1}|\Lambda|}(R)$ the potential energy curve. 
The three-photon detuning, $\Delta_{\omega_L}$, is taken with 
respect to the atomic $^2S(5s)\,\longrightarrow\, ^2P_{1/2}(6p)$ three-photon
transition. For strong photoassociation laser pulses, the 
dynamic Stark shift, $\omega_S^{^{2S+1}|\Lambda|}(t,R)$, will become
significant. It arises from the interaction of the
${}^{2S+1}|\Lambda|$ state with the intermediate off-resonant states
and is given by the effective dynamic electric dipole polarizability, 
$\alpha_{eff}(\omega_L,R)$, 
\begin{equation}
  \omega_S^{^{2S+1}|\Lambda|}(t,R)=-\frac{1}{2}|\epsilon(t)|^2\alpha_{eff}(\omega_L,R)\,,
\end{equation}
where $\epsilon(t)=|\epsilon(t)|e^{i\phi(t)}$ describes the electric
field of the laser pulse in the rotating frame
with  envelope $|\epsilon(t)|$ and $\phi(t)$ denoting 
the relative phase, taken with respect to the central frequency's phase.
$\chi^{(3)}(\omega_L,R)$ is the three-photon electric
dipole transition moment, $\xi_i(R)$ ($i=3,5$) are the spin-orbit
couplings and $A(R)$ is the non-adiabatic radial coupling matrix
element between the $(5)^1\Sigma_g^+$ and $(6)^1\Sigma_g^+$ states.  

In a second step, the initial wave packet created by the three-photon 
photoassociation, is deexcited to the
$X^1\Sigma_g^+$ ground electronic state via a  resonant
two-photon electric dipole transition. The intermediate states for the
two-photon transition are the $A^1\Sigma_u^+$ and $b^3\Pi_u$ states,
correlating to the ${}^2S(5s)+{}^2P(5p)$ asymptote, that are also
strongly coupled by spin-orbit interaction.
Electric dipole transitions are allowed between all components of the
initial wave packet and the intermediate states, whereas
the $X^1\Sigma_g^+$ ground electronic state is only connected to the
$A^1\Sigma_u^+$ state by a strong electric dipole transition.
The Hamiltonian describing the stabilization of the photoassociated
wave packet to the electronic ground state via a resonant two-photon
transition reads 
\begin{widetext}
  \begin{equation}
    \label{eq:H_dump}
    \Op{H}_{dump}(t)=\begin{pmatrix}  
      \Op{H}^{X^1\Sigma_g^+}(R) & \epsilon^*(t)d_1(R) & 0 & 0 & 0 & 0 \\
      \epsilon(t)d_1(R)  &  \Op{H}^{A^1\Sigma_u^+}(R) & \xi_1(R) & 
      \epsilon^*(t)d_2(R) & 0 & \epsilon^*(t)d_4(R)  \\
      0 & \xi_1(R) & \Op{H}^{b^3\Pi_u}(R)- \xi_2(R) & 0 & \epsilon^*(t)d_3(R) & 0   \\
      0 & \epsilon(t)d_2(R) & 0 & \Op{H}^{(5)^1\Sigma_g^+}(R) & \xi_3(R) & A(R) \\
      0 & 0 & \epsilon(t)d_3(R) & \xi_3(R) & \Op{H}^{(3)^3\Pi_g}(R) - \xi_4(R) & \xi_5(R)\\
      0 & \epsilon(t)d_4(R) & 0 & A(R) & \xi_5(R) & \Op{H}^{(6)^1\Sigma_g^+}(R)
    \end{pmatrix},
  \end{equation}
\end{widetext}
in a (one-photon) rotating wave approximation. The Hamiltonian for field-free  
nuclear motion in the ${^{2S+1}|\Lambda|}$ electronic state,
$\Op{H}^{^{2S+1}|\Lambda|}$, is now given by 
\begin{equation}
  \Op{H}^{^{2S+1}|\Lambda|}=\Op{T}+V^{^{2S+1}|\Lambda|}(R)+\Delta^{np}_{\omega_L}
\end{equation} 
with the detunings $\Delta^{5p}_{\omega_L}=\omega_{^2P(5p)}-\omega_L$ and
$\Delta^{6p}_{\omega_L}=\omega_{^2P(6p)}-2\omega_L$ for the states
dissociating into the ${}^2S(5s)+{}^2P(5p)$ and ${}^2S(5s)+{}^2P(6p)$
asymptotes, respectively. 
The electric transition dipole moments are denoted by 
\begin{eqnarray*}
d_1(R) & = & \langle X^1\Sigma_g^+| \Op d|A^1\Sigma_u^+\rangle\,,\\
d_2(R) & = & \langle A^1\Sigma_u^+| \Op d|  (5)^1\Sigma_g^+\rangle\,,\\
d_3(R) & = & \langle b^3\Pi_u | \Op d|(3)^3\Pi_g\rangle\,,\\
d_4(R) & = & \langle A^1\Sigma_u^+| \Op d|  (6)^1\Sigma_g^+\rangle\,,
\end{eqnarray*}
and the spin-orbit coupling matrix elements read
\begin{eqnarray*}
\xi_1(R)&=&\langle A^1\Sigma_u^+|\Op{H}_{SO}^{\Omega=0_u^+}| b^3\Pi_u\rangle\,,\\
\xi_2(R)&=&\langle b^3\Pi_u|\Op{H}_{SO}^{\Omega=0_u^+}| b^3\Pi_u\rangle\,,\\
\xi_3(R)&=&\langle (5)^1\Sigma_g^+|\Op{H}_{SO}^{\Omega=0_g^+}| (3)^3\Pi_g\rangle\,,\\
\xi_4(R)&=&\langle (3)^3\Pi_g|\Op{H}_{SO}^{\Omega=0_g^+}| (3)^3\Pi_g\rangle\,,\\
\xi_5(R)&=&\langle (6)^1\Sigma_g^+|\Op{H}_{SO}^{\Omega=0_g^+}| (3)^3\Pi_g\rangle\,,
\end{eqnarray*}
where $\Op{H}_{SO}$ denotes the spin-orbit Hamiltonian in the Breit-Pauli
approximation.
For large interatomic separations $R$, the transition dipole moments
and spin-orbit coupling approach their atomic values,
\begin{eqnarray*}
d_1(R\to\infty)&=&\sqrt{2}\langle ^2S(5s)|\Op d|^2P(5p) \rangle\,,\\
d_2(R\to\infty)&=&0\,,\\
d_3(R\to\infty)&=&0\,,\\
d_4(R\to\infty)&=&\sqrt{2}\langle ^2S(5s) |\Op d| ^2P(5p)\rangle\,,
\end{eqnarray*}
and
\begin{eqnarray*}
\xi_2(R\to\infty)&=&\xi_1(R\to\infty)/\sqrt{2}=\Delta_{FS}(^2P(6p))/3\,,\\
\xi_4(R\to\infty)&=&\xi_3(R\to\infty)/\sqrt{2}=\Delta_{FS}(^2P(5p))/3\,,\\
\xi_5(R\to\infty)&=&0\,,
\end{eqnarray*}
where $\Delta_{FS}(^2P(5p))=237.6\,$cm$^{-1}$ and
$\Delta_{FS}(^2P(6p))=77.5\,$cm$^{-1}$ are the atomic fine structure
splittings. 

The Hamiltonians, Eqs.~\eqref{eq:H_pump} and \eqref{eq:H_dump},
are represented on a Fourier grid with an adaptive step
size~\cite{KokooulineJCP99,WillnerJCP04,KallushCPL06} using 
$N$ = 2048 grid points. 
The time-dependent Schr\"odinger equation for the pump and dump
Hamiltonians,
\begin{equation}\label{Schrod}
  i\frac{\partial}{\partial t}|\Psi(t)\rangle=\Op{H}|\Psi(t)\rangle\,,
\end{equation}
is solved by the Chebyshev propagator~\cite{KosloffARPC94}.


\subsection{Electronic structure}
\label{subsec:elstructure}

State-of-the-art \textit{ab initio} techniques have been applied to
determine the electronic structure data of the rubidium molecule
needed in our dynamical models of the pump and dump processes. 
The details of the calculations and a thorough discussion of the
accuracy by comparison with the most recent high-resolution spectroscopic
results will be reported elsewhere \cite{TomzaJCP12}. Briefly, 
all potential energy curves for the gerade and ungerade singlet
and triplet states lying below 26000$\,$cm$^{-1}$ at the dissociation limit 
were calculated with the recently introduced Double Electron Attachment 
Intermediate Hamiltonian Fock Space Coupled Cluster method restricted to 
single and double excitations
(DEA-IH-FS-CCSD)~\cite{MusialJCP12,MusialCR12}. 
Starting with the closed-shell reference state for the doubly ionized
molecule Rb$_2^{2+}$ that shows the  correct dissociation at large distances $R$ 
into closed-shell subsystems, Rb$^+$+Rb$^+$, and using the double electron attachment 
operators in the Fock space coupled cluster ansatz renders our method 
size-consistent at any interatomic distance $R$ and guarantees the 
correct large-$R$ asymptotics. Thus, the DEA-IH-FS-CCSD approach 
overcomes the problem of the standard CCSD and equation of motion CCSD 
methods~\cite{MusialRMP07} with the proper dissociation into open-shell atoms.
The potential energy curves obtained from the \textit{ab initio}
calculations were smoothly connected at intermediate distances to 
the asymptotic multipole expansion~\cite{HeijmenMP96}. The $C_6$ coefficient of the
electronic ground state and the $C_3$ coefficient of the first excited
state were fixed at their empirical values derived from high-resolution
spectroscopic experiments~\cite{MartePRL02,GutterresPRA02}, while the
remaining coefficients were taken from Ref.~\cite{MarinescuPRA95}.

\begin{figure}[t!]
\centering
\includegraphics[width=\columnwidth]{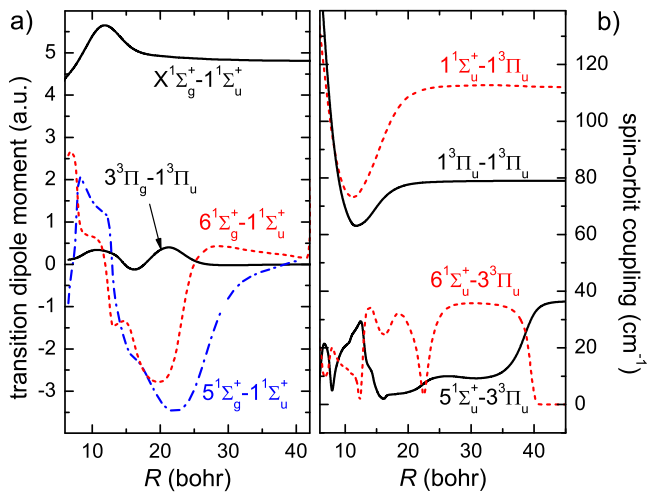}
\caption{(Color online) Electric transition dipole moments (a) and
  spin-orbit coupling matrix elements (b) essential for studied
  dynamics.} 
\label{fig:Couplings}
\end{figure}
Electric transition dipole moments, radial non-adiabatic coupling, and
spin-orbit coupling matrix elements were obtained by the multi-reference
configuration interaction method (MRCI), restricted to single and
double excitations with a large active space.  
Scalar relativistic effects were included by using the small-core fully
relativistic energy-consistent pseudopotential ECP28MDF~\cite{LimJCP05}
from the Stuttgart library. Thus, in the present study the Rb$_2$
molecule was treated as a system of effectively 20 electrons. The
$[14s14p7d6f]$ basis set was used in all calculations. This basis 
was obtained by decontracting and augmenting the basis set of 
Ref.~\cite{LimJCP05} by a set of additional functions, improving the
accuracy of the atomic excitation energies of the rubidium atom with
respect to the NIST database. With this basis set, employing the
DEA-IH-FS-CCSD method for the nonrelativistic energies and the MRCI
approach for the fine structure, we reproduce the experimental
excitation energies with a root mean square deviation (RMSD) of
21$\,$cm$^{-1}$. 
The DEA-IH-FS-CCSD calculations were done with the code based on
the \textsc{ACES II} program system~\cite{ACESII}, while the MRCI
calculations were performed with the \textsc{MOLPRO}
code~\cite{Molpro}. 

The accuracy of the computed potential energy curves is as excellent
as that of the atomic excitation energies~\cite{TomzaJCP12}.
Here, we only point out that the RMSD of the well depths of
the present potential energy curves for electronic states that were 
observed experimentally is 64$\,$cm$^{-1}$. In particular, the well
depths of the ground singlet state and the lowest triplet state
are reproduced within 2.7\% and 3.7\%, respectively. The relative
errors for the excited states relevant for the present study, the 
$A^1\Sigma_u^+$ and $b^3\Pi_u$ pair, are even smaller, 0.5\% and 0.3\%, 
respectively. We expect a similar accuracy for the higher electronic
states that have not yet been observed experimentally. 
The accuracy of the present results for the potential energy
curves is much higher than that of Refs.~\cite{ParkJMS01,MarianMP03}
and slightly better than in the recent study by Allouche and
Aubert-Fr\'econ \cite{FreconJCP12} which did not consider 
the electronic states crucial for our photoassociation proposal.


\subsection{Optimal control theory}
\label{subsec:oct}

Optimal control theory (OCT) can be used to calculate the shape of laser pulses
that efficiently drive a desired transition. We will employ it here to
determine the most efficient stabilization between an initial molecular
wave packet and deeply bound levels in the ground electronic state. In
principle, this problem is completely controllable such that perfect
population transfer can be realized. However, contraints such as limited
pulse duration, spectral bandwidth and pulse intensity will compromise
the stabilization process, reducing the transfer efficiency. 

The control problem is defined by minimization of the functional
\begin{equation}\label{J}
  J=J_T+\int_0^T g[\epsilon(t)]dt\,,
\end{equation}
where the first term denotes the final-time $T$ target and the second
one intermediate-time costs. 
The final-time target, $J_T$, can be chosen to correspond to a single
state-to-state transition, $J_T^{ss}$, or to the transition into a
manifold of final states, $J_T^{sm}$. For a single state-to-state
transition from an initial state $|\Psi_{in}\rangle$ to a target
state, here a vibrational level of the electronic ground state,
$v''$, the final-time functional is written as 
\begin{equation}\label{eq:Jss}
  J^{ss}_T=1-|\langle\Psi_{v''}|\Op{U}(T,0;\epsilon) |\Psi_{in} \rangle|^2\,.
\end{equation}
$\Op{U}(T,0;\epsilon) |\Psi_{in} \rangle$ represents the formal
solution of the time-dependent Schr\"odinger equation with
$\Op{U}(T,0;\epsilon)$ the time evolution operator. 
$J_T^{ss}$ corresponds the overlap of the initial state, propagated
to the final time $T$ under the action of the laser field
$\epsilon(t)$, with the target state. Optimizing a transition into a
manifold of states is expressed by the functional
\begin{equation}\label{eq:Jsm}
  J^{sm}_T=1-\sum_{v''=v''_{min}}^{v''_{max}}|
  \langle\Psi_{v''}|\Op{U}(T,0;\epsilon)|\Psi_{in}\rangle|^2 \,, 
\end{equation}
where any vibrational level of the electronic ground state
with quantum number between $v''_{min}$ and $v''_{max}$  can be
populated at the final time. Once the optimum is reached, 
both functionals, $J_T^{ss}$ and $J_T^{sm}$, take the value zero.

The intermediate time cost, $g[\epsilon(t)]$, can in general depend on
both the state and the field. Here we restrict the dependence to one
on the laser field only, where we ask that optimization does not
change, or changes only minimally, the integrated pulse energy,
\begin{equation}
  g[\epsilon(t)]=\frac{\lambda}{S(t)}\sum_{a=\{\mathfrak{Re,Im}\}}
  \left(\epsilon_a^{(k+1)}(t)-\epsilon_a^{(k)}(t)\right)^2\,,
\end{equation}
with $k$ labelling the iteration step~\cite{PalaoPRA03}. The shape
function $S(t)$, $S(t)=\sin^2(\pi t/T)$, 
enforces a smooth switch on and off of the field and $\lambda$ is a
weight. Note that the laser field is complex since we employ the
rotating-wave approximation. A non-zero phase indicates a relative
phase with respect to the laser pulse peak center, or, in the spectral
domaine, with respect to the phase of the central laser frequency. 

Using the linear variant of Krotov's
method~\cite{SklarzPRA02,ReichJCP12}, the update equation for the
laser field at iteration step $k+1$ can be derived,
\begin{widetext}
  \begin{eqnarray}\label{eq:update}
    \epsilon_{\mathfrak{Re/Im}}^{(k+1)}(t)
    &=&\epsilon_{\mathfrak{Re/Im}}^{(k)}(t) -
    \frac{S(t)}{2\lambda}\mathfrak{Im}\Bigg\{
      \sum_{v''=v''_{min}}^{v''_{max}}\left\langle \Psi_{in}\left|
      \Op{U}^\dagger\left(T,0;\epsilon^{(k)}\right)\right|\Psi_{v''}\right\rangle 
    \\ && \nonumber
    \quad\quad\quad\quad\quad\quad\quad\quad\quad\quad\quad\quad
    \quad
      \left\langle\Psi_{v''} \left|\Op{U}^\dagger\left(t,T;\epsilon^{(k)}\right) 
      \frac{\partial\Op{H}_{dump}}{\partial\epsilon_{\mathfrak{Re/Im}}}
    \bigg|_{\epsilon_{\mathfrak{Re/Im}}^{(k+1)}}
      \Op{U}\left(t,0;\epsilon^{(k+1)}\right)\right|\Psi_{in}\right\rangle\Bigg\},
  \end{eqnarray}
\end{widetext}
where $\Op{U}(t,0;\epsilon^{(k+1)})|\Psi_{in}\rangle$ is the initial
state forward propagated  to time $t$ under the action of the new
field, $\epsilon^{(k+1)}$, and 
$\Op{U}(t,T;\epsilon^{(k)})|\Psi_{v''}\rangle$ denotes the target
state(s) backward propagated  to time $t$ under the action of the old
field, $\epsilon^{(k)}$.
The derivative of the Hamiltonian with respect to
the field yields a matrix having as its elements all the
transition dipole moments $d_i$, cf. Eq.~\eqref{eq:H_dump}. 
Optimization of the functionals $J_T^{ss}$ or $J_T^{sm}$ requires repeated
forward and backward propagations of the initial and target states.


\section{Excited state wave packet representing the initial 
  state for stabilization} 
\label{sec:ini}

The initial wave packet for the stabilization step is created by the
photoassociation pulse. 
The simplest pulse that can be employed for the  three-photon
photoassociation is a transform-limited (TL) Gaussian pulse. 
The intensity of the laser pulse is chosen to be in the perturbative
week-field regime, where the composition of the photoassociated wave
packet reflects the bandwidth of the laser pulse combined with the
vibrationally averaged three-photon electric dipole transition moments
between the initial scattering state and the excited state vibrational
levels below ${}^2S(5s)+{}^2P_{1/2}(6p)$ dissociation limit. 
A pulse duration of 4$\,$ps full width at half maximum (FWHM) is
considered, corresponding to a spectral bandwidth of
3.7$\,$cm$^{-1}$. The pulse is red detuned by 12$\,$cm$^{-1}$
from the $^2S(5s)\,\longrightarrow\,^2P_{1/2}(6p)$ atomic three-photon
transition.   
In order to utilize broadband femtosecond laser pulses, 
more elaborate pulse shapes will be required that suppress the
excitation of atoms~\cite{KochFaraday09} while possibly maximizing
free-to-bound transitions. However, the general features of the
photoassociated wave packet are determined by the three-photon matrix
elements. They are the largest close to the ${}^2S(5s)+{}^2P_{1/2}(6p)$
dissociation limit, corresponding to a photoassociation window
at interatomic separations between 30$\,a_0$ and 50$\,a_0$.
Transition moments to the vibrational levels detuned by
more then 30$\,$cm$^{-1}$ from the ${}^2S(5s)+{}^2P_{1/2}(6p)$
threshold, corresponding to photoassociation windows at shorter
interatomic separations, are significantly smaller.   
The most important contributions to the
photoassociated wave packet will therefore remain the same as in our
simple example. 
It could turn out that, using coherent control, photoassociation into
wave packets with binding energies larger than 30$\,$cm$^{-1}$ becomes
feasible. In this case, the stabilization of the photoassociated
molecules becomes easier, and their faster vibrational dynamics 
and larger Franck-Condon factors to the deeply bound
$X^1\Sigma_g^+$ vibrational levels will only improve the predictions
of the present study. 

\begin{figure}[t!]
\centering
\includegraphics[width=\columnwidth]{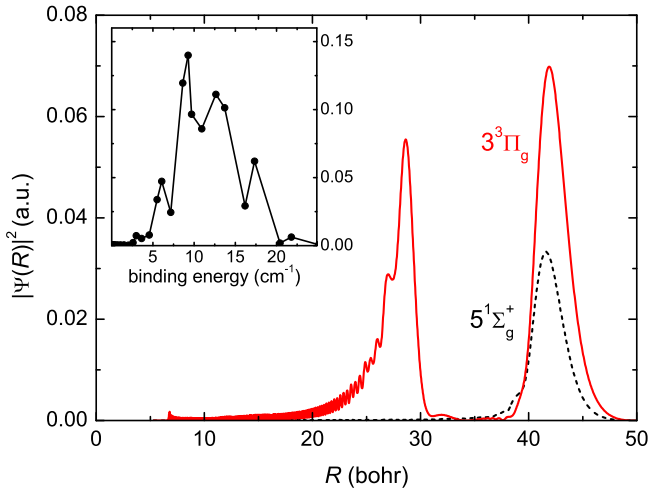}
\caption{(Color online) Wave packet obtained after photoassociation
  with a transform-limited pulse. 
  The pulse duration is 4$\,$ps (FWHM) and a snapshot 2.5$\,$ps after
  the photoassociation pulse
  maximum is shown. At this time, the $(6)^1\Sigma_g^+$ component is
  insignificant and not visible on the scale of this figure.
  Inset: Decomposition of the wave packet 
  onto the vibrational levels of the coupled $(5)^1\Sigma_g^+$,
  $(6)^1\Sigma_g^+$, and $(3)^3\Pi_g$ states.} 
\label{fig:Wavepacket}
\end{figure}
The initial wave packet for the stabilization step 
is plotted in Fig.~\ref{fig:Wavepacket}.
The inset shows the decomposition of the wave packet 
onto the vibrational levels of the
coupled $(5)^1\Sigma_g^+$,  $(6)^1\Sigma_g^+$, and $(3)^3\Pi_g$
states. The binding energy of the wave packet amounts to
11.55$\,$cm${}^{-1}$. The snapshot shown in Fig.~\ref{fig:Wavepacket}
is taken 2.5$\,$ps after the maximum of the photoassociation pulse,
i.e., before the pulse is over. At this time, the Gaussian character
of the wave packet is still apparant, while at later times the
strongly anharmonic shape of the potential energy curves leads to
strong wave packet dispersion. Note that the photoassociated wave
packet shows truly mixed character with about 65\% of its norm
residing on the $(3)^3\Pi_g$ triplet component and 35\% on  the
$(5)^1\Sigma_g^+$ singlet component. This is despite the fact that 
electric dipole transitions are allowed 
only between the atomic pair in the  $a^3\Sigma_u^+$ lowest triplet
state and the $(3)^3\Pi_g$ triplet component of the coupled
electronically excited manifold dissociating into
${}^2S(5s)+{}^2P_{1/2}(6p)$ and illustrates the strong spin-orbit
coupling. The role of the spin-orbit coupling is further evidenced by
the double peak structure of 
the wave packet with the short-range peak corresponding to the outer
turning point of the $(3)^3\Pi_g$ potential and the long-range peak
reflecting the outer turning point of the $(5)^1\Sigma_g^+$ potential. 
The importance of the strong resonant spin-orbit coupling for the
stabilization of photoassociated molecules has been discussed
extensively~\cite{DionPRL01,PechkisPRA07,FiorettiJPB07,GhosalNJP09,LondonoPRA09,TomzaPCCP11,SkomorowskiPRA12}.
In the present study, we will not only use it for improved 
transition matrix elements to deeply bound ground state levels, but
also to convert an atom pair colliding in the triplet state to singlet
molecules~\cite{SagePRL05}.  

\begin{figure}[t!]
\centering
\includegraphics[width=\columnwidth]{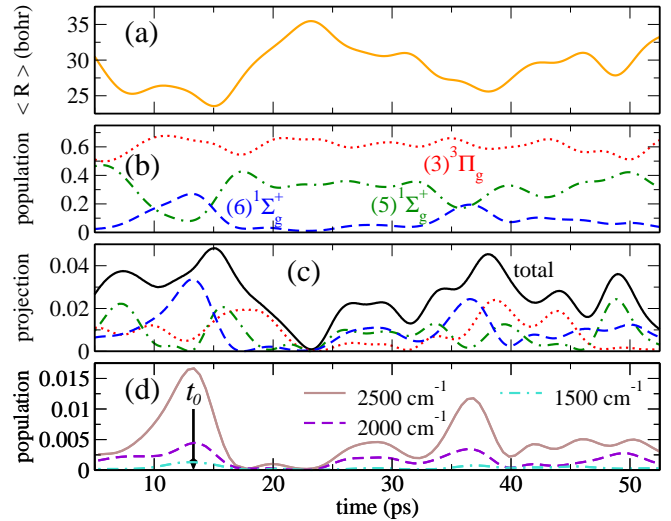}
\caption{(Color online) a: Time evolution of the average bond
  length of the wave packet. b: Time evolution of the populations
  of the $(5)^1\Sigma_g^+$, $(6)^1\Sigma_g^+$, and $(3)^3\Pi_g$ components
  of the wave packet. 
  c: Projection
  of the time-dependent wave packet, and its $(5)^1\Sigma_g^+$,
  $(6)^1\Sigma_g^+$, and $(3)^3\Pi_g$ components, onto all vibrational
  levels of the $X^1\Sigma_g^+$ ground electronic states with binding
  energies up to 1000$\,$cm$^{-1}$. 
  d: Population of the $X^1\Sigma_g^+$ component of the
  wave function after stabilization with a TL pulse 
  vs time delay between
  photoassociation and stabilization pulse for three 
  different detunings, taken with the respect to the 
  $^2P_{1/2}(6p)\,\longrightarrow\,^2S(5s)$ two-photon transition.
  The stabilization pulse has a FHWM of 100$\,$fs and an integrated
  pulse energy of 25.4$\,$nJ  corresponding to the weak field regime. 
  The arrow indicates $t_0=13.3\,$ps chosen as the time delay between
  the photoassociation and stabilization pulse peaks.
} 
\label{fig:Surfnorms}
\end{figure}
The initial wavepacket propagates toward shorter interatomic
separations under the influence of the excited state potentials.
At large interatomic separations, the potential energy curve of the
$(5)^1\Sigma_g^+$ state displays a strong $-1/R$ ion-pair character. 
The singlet-triplet oscillations are analyzed in
Fig.~\ref{fig:Surfnorms}(a) displaying the singlet and
triplet components of the wave packet evolving after the
photoassociation pulse in the manifold of electronically excited
states. 
The population of the $(3)^3\Pi_g$ triplet component oscillates
around 60\%, whereas the population of the $(6)^1\Sigma_g^+$ component,
that was absent just after photoassociation, reaches a maximum of 27\% 
at $t=12.7\,$ps after the peak of photoassociating pulse. A second
maximum of the $(6)^1\Sigma_g^+$ component is observed after a period of 
20.1$\,$ps and a third one after another 36.2$\,$ps. The times at
which the $(6)^1\Sigma_g^+$ component reaches maximal values can
be interpreted as moments when the wave packet arrives at its shortest
distance and is reflected from the innermost repulsive short range wall. 
This observation is confirmed by calculating the average bond length of the
wave packet, shown in Fig.~\ref{fig:Surfnorms}(b),
which allows to estimate 
the revival time of the present wave packet to be between 20$\,$ps
and 30$\,$ps. 
This estimate agrees with the range of revival times, defined by
$T_{rev}(v)=\frac{2h}{|E_{v+1}+E_{v-1}-2E_v|}$~\cite{AverbukhPLA89}, 
for the vibrational levels $v$ that make up a wave packet with binding
energies close to 12$\,$cm$^{-1}$.   

The knowledge of the revival time of the wave packet is useful for the
interpretation of the projections of the time-dependent wave packet,
$|\langle \Psi_{in}(t)|v''\rangle|^2$, onto the 
vibrational levels $v''$ of the $X^1\Sigma_g^+$ ground electronic state,
shown in Fig.~\ref{fig:Surfnorms}(c) for all ground
state levels with binding energies up to 
1000$\,$cm$^{-1}$. These projections are largest when the wave packet
is localized at its inner turning point, cf. 
Fig.~\ref{fig:Surfnorms}(b). The time at which the projections show
maxima correspond to optimal time delays between photoassociation and  
stabilization pulse. The times inbetween these maxima are given by the  
revival time. However, the transition probability does not only depend
on the overlap of initial and final wave function, but also on the
dipole moments and the topology of the intermediate state surfaces and
their coupling. This is illustrated by the difference between 
Fig.~\ref{fig:Surfnorms}(c) and (d) with (b)
showing the calculated population on the $X^1\Sigma_g^+$ ground
electronic state as a function of
the time delay between photoassociation and stabilization pulse
for three different detunings of the stabilization pulse.
The $X^1\Sigma_g^+$ population is obtained by solving the time-dependent
Schr\"odinger equation for a weak TL stabilization pulse. 
The positions of the maxima and minima of the final $X^1\Sigma_g^+$
state population correspond to those of the projection
of the $(6)^1\Sigma_g^+$ component (dashed blue line in 
Fig.~\ref{fig:Surfnorms}(c)) rather than the total projection (black
line in Fig.~\ref{fig:Surfnorms}(c)). This suggests that the transition 
from the $(6)^1\Sigma_g^+$ component is the most important one 
in the stabilization process. 

Based on the time-dependence of the projection and the stabilization
probability analyzed in Fig.~\ref{fig:Surfnorms}, we choose
$t_0=13.3\,$ps for the time delay, taken between the peak of the
photoassociating pulse and the center of all pulses used in the
following sections.


\section{Stabilization to the electronic ground state with 
  transform-limited and linearly chirped
  pulses} 
\label{sec:anapulse} 
We first study transform-limited and linearly chirped stabilization
pulses, in order to understand the role of the basic pulse
parameters such as intensity and spectral width and to investigate
dynamical effects. By comparing projections and actual final state
populations in the previous section, we have shown that a
simple two-photon Franck-Condon principle does not correctly capture
the stabilization dynamics. A less simplified picture is obtained by
taking the structure of the vibrational levels in the initial,
intermediate and final electronic states fully into account,
neglecting strong field effects and a dynamical interplay between
pulse and spin-orbit couplings. Specifically, in the weak-field regime
and for TL pulses, the probability of the resonant two-photon
transition is obtained by perturbation theory. It is determined by the
effective two-photon transition moment, 
\begin{equation}\label{eq:Twophoton}
  D(v'')=\sum_{v'}{\left|
\sum_{j=2,3,4}
\langle \Psi_{in}|d_j|v'\rangle
 \langle v'|d_1|v''\rangle\right|} e^{\Delta\omega_{v'}^2/2\sigma_\omega^2} \,,
\end{equation}
where $\langle \Psi_{in}|d_j|v'\rangle=
\sum_{i,i'}\int\Psi_{in}^i(R)^*d_j(R)\chi_{v'}^{i'}(R)dR$ denotes the 
electric transition dipole moment between the initial wave packet with
components $i$ in the $(5)^1\Sigma_g^+$, $(6)^1\Sigma_g^+$, and
$(3)^3\Pi_g$ manifold and the intermediate vibrational levels 
$v'$ with components $i'$ in the excited $A^1\Sigma_u^+$ and
$b^3\Pi_g$.  
Correspondingly, $\langle v'|d_1|v''\rangle$ is the electric 
transition dipole moment between the intermediate levels $v'$ of the
$A^1\Sigma_u^+$ and $b^3\Pi_g$ states and the final ground $X^1\Sigma_g^+$ 
state vibrational level $v''$. The exponent accounts for the bandwidth of
the laser pulse, $\sigma_\omega$, and the detuning of the intermediate
levels from the laser frequency,
\begin{equation}
  \Delta\omega_{v'}=\omega_{^2P_{1/2}(5p)}-\omega_{v'}+\omega_{v''}-\omega_L\,.
\end{equation}
The laser frequency, $\omega_L$, is chosen such that the two-photon
transition is resonant for the initial wave packet and the final level
$v''$, $\omega_L=(\omega_{^2P_{1/2}(6p)}-\omega_{in}+\omega_{v''})/2$. 
Here, $\omega_{in}$ denotes the binding energy of the initial wave
packet, 
defined to be positive, and $\omega_{{}^2P_{1/2}(np)}$ is the
excitation energy of the ${}^2P_{1/2}(np)$ atomic level. The spectral
width is given in terms of the FWHM of the time profile of the pulse,
$\tau$, $\sigma_\omega=2\sqrt{2\ln(2)}/\tau$. 

\begin{figure}[t!]
\centering
\includegraphics[width=\columnwidth]{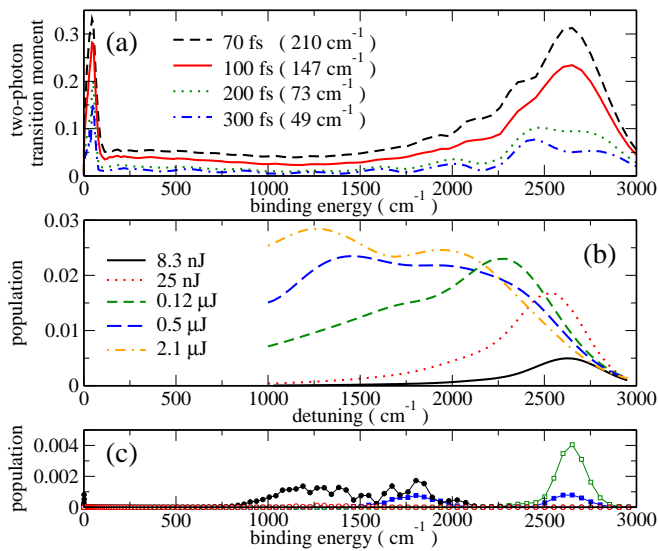}
\caption{(Color online) a: Resonant two-photon transition moments defined 
  by Eq.~\eqref{eq:Twophoton} vs binding energy of the target 
  $X^1\Sigma_g^+$ state levels $v''$
  for four different pulse durations or spectral widths,
  respectively. 
  b: Final population of the
  $X^1\Sigma_g^+$ state after a TL 100$\,$fs pulse 
  vs pulse detuning, taken with the respect to the
  $^2P_{1/2}(6p)\,\longrightarrow\,^2S(5s)$ atomic two-photon
  transition, for five different integrated pulse energies.
  c: Vibrational decomposition of the $X^1\Sigma_g^+$ component 
  of the wave function after stabilization with TL 100$\,$fs pulses with 
  different pulse energies and detunings with the respect to 
  $^2P_{1/2}(6p)\longrightarrow{}^2S(5s)$ atomic two-photon transition: 
  $\square$~--~25.4$\,$nJ and 2650$\,$cm$^{-1}$, $\blacksquare$~--~2.06$\,\mu$J
  and 2650$\,$cm$^{-1}$, $\Circle$~--~25.4$\,$nJ and 1240$\,$cm$^{-1}$, 
  $\CIRCLE$~--~2.06$\,\mu$J and 1240$\,$cm$^{-1}$.
} 
\label{fig:Twophoton}
\end{figure}
The two-photon transition moments, defined by Eq.~\ref{eq:Twophoton},
are shown in Fig.~\ref{fig:Twophoton} for all vibrational levels $v''$
of the $X^1\Sigma_g^+$ ground electronic state. The large peak around
a binding energy of 2650$\,$cm$^{-1}$ indicates that stabilization of
the excited state wave packet to levels with binding energies in this
range  is most efficient. The peak maximum in Fig.~\ref{fig:Twophoton}
corresponds to a transition to the vibrational level $v''=23$, with
a binding energy $E_{v''}$=2651$\,$cm$^{-1}$. This suggests that the
level $v''=23$ might be a good choice for the target in
the state-to-state optimization of stabilization process below in
Section~\ref{sec:oct}. 
A standard choice of 100$\,$fs pulse duration for the TL pulse,
corresponding to a spectral width of about 150$\,$cm$^{-1}$ FWHM is
sufficient to address a broad distribution of target vibrational
levels in the $X^1\Sigma_g^+$ state.

The integrated pulse energy is given by
\begin{equation}
E=\varepsilon_0 cA\int_0^\infty \left|\epsilon(t)\right|^2dt\,,
\end{equation}
with $\epsilon(t)$ the laser field, $A=\pi r^2$ the area which is covered 
by the laser ($r=50\,\mu$m was assumed), $c$ the speed of light, and $\varepsilon_0$ 
the dielectric constant. 
We use the integrated pulse energy rather than the peak
intensity of the pulse since, independently of the pulse duration, it
quantifies the energy pumped into the molecule.

The two-photon transition probability can be predicted from the
effective two-photon transition moment, cf. Eq.~\eqref{eq:Twophoton}
and Fig.~\ref{fig:Twophoton}, only in the weak-field regime
when dynamic Stark shifts and other time-dependent effects do
not play any role. The dependence of the  two-photon transition
probability on the pulse intensity and detuning,
\begin{equation}
  \label{eq:detuning}
  \Delta\omega_L = 2\omega_L-\omega_{{}^2P_{1/2}(6p)}
\end{equation}
 is illustrated in 
Fig.~\ref{fig:Twophoton}(b). The pulse duration is kept
fixed at 100$\,$fs FWHM. 
For weak and intermediate pulse intensities, with the integrated pulse
energy corresponding to 8.3$\,$nJ and 25.4$\,$nJ, the 
final $X^1\Sigma_g^+$ ground state population as a function of the
pulse detuning reflects the shape of the effective two-photon
transition moment, Fig.~\ref{fig:Twophoton}(a). On the other hand, 
the final $X^1\Sigma_g^+$ population decreases for the detuning
corresponding to the maximum of the two-photon transition moment and
increases for smaller detunings when the integrated pulse energy is
increased. This observation is rationalized in terms of the 
strong dynamic Stark shift by analyzing the vibrational
distribution of the final $X^1\Sigma_g^+$ state population in 
Fig.~\ref{fig:Twophoton}(c) for a detuning, 
$\Delta\omega_L$=2650$\,$cm$^{-1}$, corresponding to the maximum of
the two-photon transition moment in Fig.~\ref{fig:Twophoton}(a). 
When increasing the integrated pulse energy from 25.4$\,$nJ to
2$\,\mu$J, i.e., from the intermediate to the strong
field regime, two peaks are observed in the vibrational distribution
rather than a single Gaussian around the binding energy of $v''=23$, 
reflecting the bandwidth of the pulse. In the strong field regime, one
peak of the vibrational distribution is still located around the
binding energy of the resonant level, while the second one is shifted
by 800$\,$cm$^{-1}$ to smaller binding energies. 
This is due to the positive differential Stark shift caused by the
coupling to the intermediate states,
which indeed increases the energy separation between ground and
excited states by about 800$\,$cm$^{-1}$. 
The dynamic Stark shift of the ground state vibrational level $v''$
is estimated according to 
\begin{equation}
  \omega_S^{v''}=-\frac{1}{2}|\epsilon(t)|^2
  \sum_{v''}|\langle v''|\Op{d_e}|v'\rangle|^2
  \frac{\omega_{v'v''}}{\omega_{v'v''}^2-\omega_L^2}\,, 
\end{equation}
where $\langle v''|\Op{d_e}|v'\rangle$ is the electric transition
dipole moment and $\omega_{v'v''}$ the
transition frequency between levels $v'$ and $v''$, and $\omega_L$
denotes the laser frequency. 
Figure~\ref{fig:Twophoton}(c) compares the final
state vibrational distribution for two different detunings of the
stabilization pulse, $\Delta\omega_L=2650\,$cm$^{-1}$ 
corresponding to the peak of the two-photon transition probability for
weak and intermediate fields (black dotted and red solid curves in 
Fig.~\ref{fig:Twophoton}(b)) and
$\Delta\omega_L=1240\,$cm$^{-1}$  corresponding to the peak of the
two-photon transition probability for strong fields (black dot-dashed
curve in Fig.~\ref{fig:Twophoton}(b)). Inspection
of the vibrational distributions in 
Fig.~\ref{fig:Twophoton}(c) reveals that for
$\Delta\omega_L=1240\,$cm$^{-1}$ and $2\,\mu$J integrated pulse
energy, a peak at binding energies larger 
than the detuning appears. Also, this peak is caused by the
differential Stark shift which this time is negative, decreasing the
energy separation between ground and excited states by about
600$\,$cm$^{-1}$. In the weak and intermediate field regime, almost no
population is transferred for $\Delta\omega_L=1240\,$cm$^{-1}$ (red
empty circles in Fig.~\ref{fig:Twophoton}(c)),
confirming a strong field effect. 

The total population that is transferred by a TL pulse,
with 100$\,$fs FWHM and a detuning in the range of 2500$\,$cm$^{-1}$
to 2600$\,$cm$^{-1}$, from
the initial wave packet to the $X^1\Sigma_g^+$ ground electronic state
amounts to up to 1.7\% in the weak and intermediate field regime.
For strong fields, up to 2.9\% of the population can be transferred
for detunings between 1000$\,$cm$^{-1}$  and 1500$\,$cm$^{-1}$ and
pulse energies above 1$\,\mu$J. The subsequent analysis is restricted 
to pulses with detunings between 2500$\,$cm$^{-1}$ and
2650$\,$cm$^{-1}$ corresponding to the maximum of the effective 
two-photon transition moment where the smallest pulse intensities
should be required. 

\begin{figure}[t!]
  \centering
  \includegraphics[width=\columnwidth]{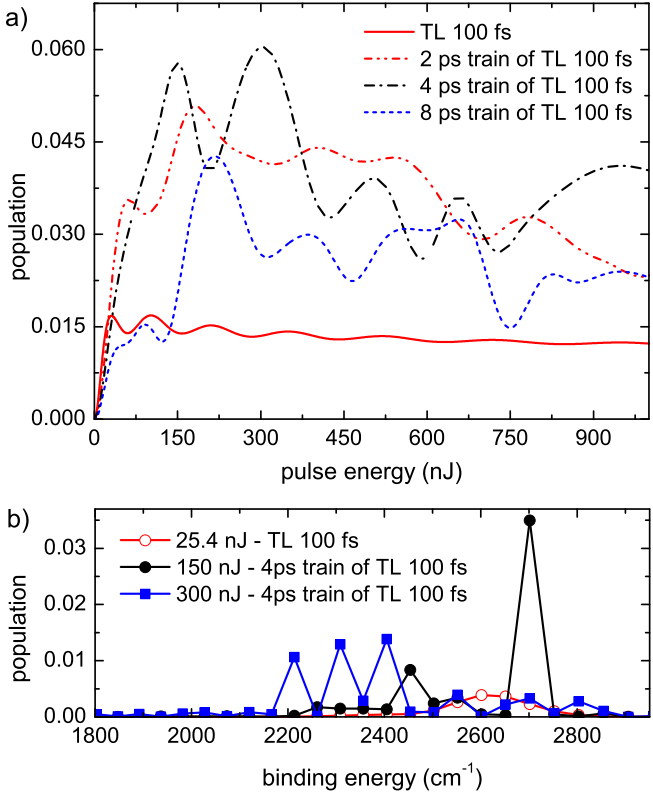}
  \caption{(Color online) a: 
    $X^1\Sigma_g^+$ state population after a  100$\,$fs TL
    pulse (solid line) and a train of 100$\,$fs TL pulses (dashed and
    dotted lines) vs integrated pulse  energy. 
    b: $X^1\Sigma_g^+$ state  vibrational distribution
    after stabilization with a 100$\,$fs TL
    pulse and 4$\,$ps train of 100$\,$fs TL pulses.
    The pulse detuning is 2500$\,$cm$^{-1}$ taken with the respect to
    $^2P_{1/2}(6p)\,\longrightarrow\,^2S(5s)$ atomic two-photon transition. 
  }
  \label{fig:TLvsEnergy}
\end{figure}
As apparant from Fig.~\ref{fig:Twophoton}, the dynamic Stark shift
is detrimental to efficient population transfer by the stabilization
pulse. One option to increase the integrated pulse power while keeping
the maximum field intensity, and thus the dynamic Stark shift, 
small is to consider a train of short TL
pulses. A second option is given by chirping the pulse. The efficiency
of the two-photon population transfer to the electronic ground with
the first option, a train of 100$\,$fs TL pulses delayed relative to
each other by 200$\,$fs and with a sinusoidal envelope,
is analyzed in Fig.~\ref{fig:TLvsEnergy}. 
While increasing the pulse energy of a 100$\,$fs pulse
does not improve the population transfer to the electronic ground
state beyond 1.7\%, a train of pulses yields up to about 6\% for pulse
energies that are still in the nJ range. The population transfer with 
a train of short pulses can be interpreted as the cumulative result
of many single transitions that accumulate amplitude in the
$X^1\Sigma_g^+$ ground state. Using a train of pulses instead of a
single pulse with the same bandwidth, one can produce 3.5 times more
ground state molecules.  Fig.~\ref{fig:TLvsEnergy}
thus confirms that the Stark shift is responsible for the
comparatively inefficient population transfer observed for TL pulses. 

\begin{figure}[t!]
  \centering
  \includegraphics[width=\columnwidth]{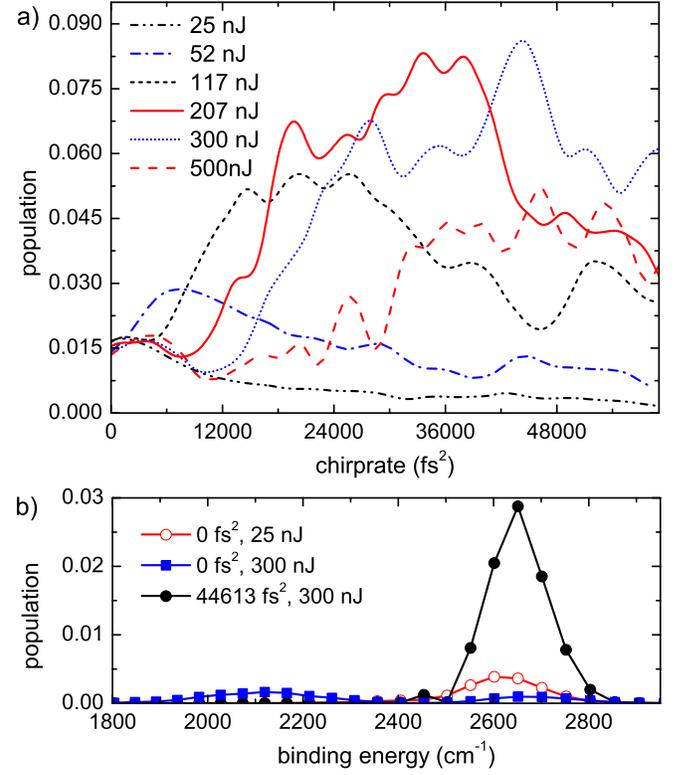}
  \caption{(Color online)  a: Final 
    $X^1\Sigma_g^+$ state population after linearly chirped pulses
    with a TL FWHM of 100$\,$fs
    vs chirp rate for different pulse energies. 
    b:   $X^1\Sigma_g^+$ state  vibrational distribution
    after stabilization with a weak TL pulse, a TL and a
    chirped strong pulse.
    The pulse detuning is 2500$\,$cm$^{-1}$ taken with the respect to
    $^2P_{1/2}(6p)\,\longrightarrow\,^2S(5s)$ atomic two-photon transition. 
}
\label{fig:Chirped}
\end{figure}
From the coherent control of atomic transitions using strong fields,
it is known that the influence of the dynamic Stark shift can be
compensated by chirping the pulse~\cite{TralleroPRA05,WeinachtPRL06}. 
We investigate in Fig.~\ref{fig:Chirped}(a) showing
the final ground state population vs chirp rate for increasing pulse
energy whether this approach works also for molecular transitions.
We use a positive chirp to correct the influence of the dynamic Stark
shift since the differential Stark shift for stabilization to
vibrational levels with binding energies close to 2650$\,$cm$^{-1}$ is
positive.   
Chirping a weak-field pulse (black dot-dashed curve) deteriorates the
population transfer. When more energetic pulses are used, chirping
increases the final  $X^1\Sigma_g^+$ state population from 1.5\%
for unchirped pulses to almost 9\% for the best chirped pulses.
In total we find that chirping the pulse improves the 
stabilization process and enhances the amount of ground state
population by a factor of about six. 
Figure~\ref{fig:Chirped}(b) showing the final state
vibrational distribution confirms that the same mechanism as in the
atomic case is at work~\cite{TralleroPRA05,WeinachtPRL06}:
When increasing the pulse energy from 25.4$\,$nJ to 300$\,$nJ without
chirping the pulse, a second peak shifted by 500$\,$cm$^{-1}$
appears. The energies of the levels of this second peak correspond
exactly to the detuning corrected by the Stark shift.
A linear chirp introduces a time-dependent instantaneous frequency of
the pulse, $\omega(t)=\omega_L+\chi t/2$, with $\chi$ the temporal
chirp rate. When chosen correctly, the chirp compensates the phase
that the molecules accumulate due the Stark shift and thus  prevents
the transition to shift out of resonance. This leads to the 
strong enhancement of the stabilization efficiency observed in
Fig.~\ref{fig:Chirped}. 

Our investigation of the stabilization dynamics under TL
and linearly chirped pulses 
shows that simply replacing a strong TL pulse by a train of pulses
with the same total integrated pulse energy or 
linearly chirping the pulse can enhance the
stabilization probability from 1\% up to 9\%. The reason for the
enhancement is given by the weaker Stark shifts for smaller peak
intensities and compensation of the phase accumulated due to the Stark
shift by a linear chirp. In the following section we will employ
optimal control theory to calculate the optimum detunings and pulses
shapes. This will allow us to determine the maximum number of ground
state  molecules that can be produced for a given integrated pulse
energy. 


\section{Stabilization to the electronic ground state 
  with optimally shaped pulses}
\label{sec:oct}

\begin{figure}[t!]
\centering
\includegraphics[width=\columnwidth]{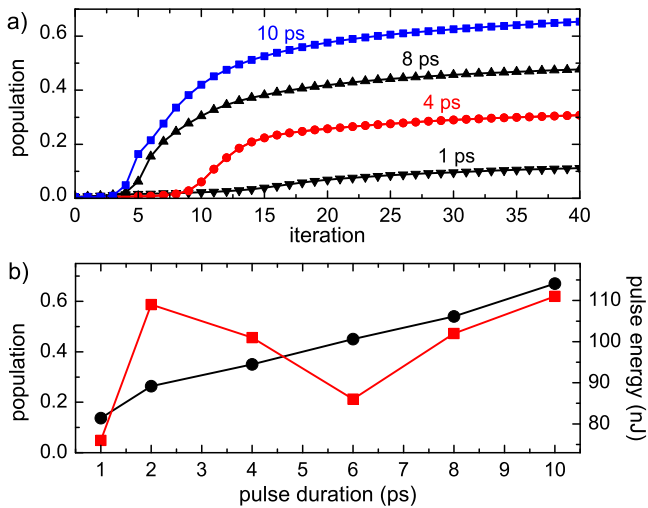}
\caption{(Color online) a: Final $X^1\Sigma_g^+$ state population
  after optimized pulses of different
  pulse duration vs number of iteration steps. 
  b: Final $X^1\Sigma_g^+$ state  population 
  (left scale, black circles) and corresponding pulse energy
  (right scale, red squares) vs pulse duration. The
  values were collected after 50 iterations, starting with the same
  weak-field guess pulse ($E=10\,$nJ) and using the same weight,
  $\lambda=400$, in all optimizations.  
}
\label{fig:OCT_conv}
\end{figure}
We employ the optimization algorithm described in
Sec.~\ref{subsec:oct} to find those laser pulses that stabilize the
initial wave packet most efficiently to the $X^1\Sigma_g^+$ electronic 
state. The final $X^1\Sigma^+_g$ state population, shown in
Fig.~\ref{fig:OCT_conv}(a),  converges
smoothly to the maximal value that can be obtained with a given pulse
duration, displayed in Fig.~\ref{fig:OCT_conv}(b). The maximum 
stabilization probability for a pulse duration of 1$\,$ps is
14\%. Increasing the pulse duration, the stabilization probability
reaches 26\% for 2$\,$ps pulse and 67\% for 10$\,$ps pulses. 
The integrated pulse energies of the optimized pulses vary between 80$\,$nJ
and 150$\,$nJ. This is two to three times smaller than the integrated
pulse energies for the trains of pulses and the linearly chirped
pulses discussed in Sec.~\ref{sec:anapulse}. The guess pulse for the
optimizations shown in Fig.~\ref{fig:OCT_conv} is a TL pulse with a
pulse duration of 100$\,$fs and integrated pulse energy of 
10$\,$nJ. For all the results presented here, the 
state-to-manifold-of-states functional, Eq.~\eqref{eq:Jsm}, was
employed. The results obtained by using the state-to-state functional,
Eq.~\eqref{eq:Jss}, do not differ significantly. In particular, the same
bounds on the maximum stabilization efficiencies are
observed. However, for the state-to-state functional, the
optimizations were found to converge slower. This is easily
rationalized in terms of a single state being a more restrictive
optimization target than a manifold. The integrated energy
of the optimal pulses presented in Fig.~\ref{fig:OCT_conv}(b) does not
depend strongly on the optimal pulse duration. The slightly oscillatory
behavior of the integrated pulse energy as a function of pulse
duration is observed irrespective of the shape and energy of the guess
pulse and the weight $\lambda$. 

The fact that the maximum population transfer to the $X^1\Sigma_g^+$
state is clearly less than 100\% is due to the pulses being too short
to drive the complete wave packet to the ground state~\cite{KochPRA04}. 
When the pulse
duration is much shorter than the time scale of the vibrational motion
and spin-orbit oscillations, then only that part of the wave packet
that shows a favorable overlap with the target state during the
optimization window is transferred. For example, the pulse with 
1$\,$ps duration essentially reflects the overlap of the initial wave
packet.  By increasing the pulse duration,
cf. Fig.~\ref{fig:OCT_conv}(b), the stabilization efficiency increases
monotonically. A stabilization probability
of 100\% is expected once the pulse duration is longer than 
the revival time of the wave packet, estimated above to be
between 20$\,$ps and 30$\,$ps. In fact, optimizations with pulse
durations of 20$\,$ps and 30$\,$ps yield stabilization efficiencies of
93\% and 99\%, respectively, with pulse energies below
150$\,$nJ. However, we restrict our analysis
to  pulse durations up to 10$\,$ps since stretching a TL 100$\,$fs
pulse by more than a factor of 100 due to the pulse shaping is not
realistic. 

\begin{figure}[t!]
\centering
\includegraphics[width=\columnwidth]{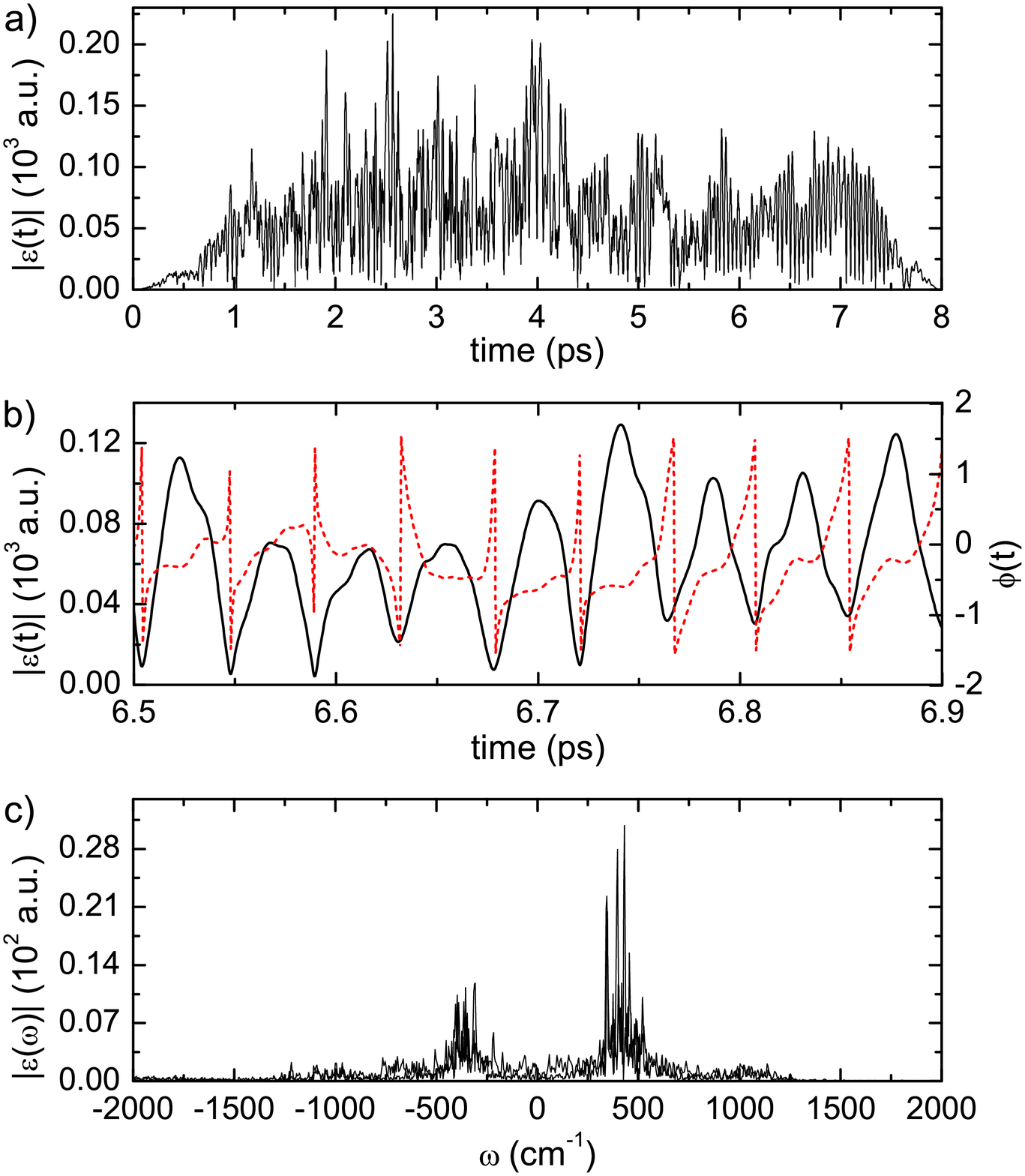}
\caption{(Color online) Temporal envelope (a) and spectrum (c) of an
  optimized pulse. 
  b: Envelope (left scale, solid black line) 
  and temporal phase (right scale, red dashed line) of the optimized
  pulse in a short time interval.
}
\label{fig:OCT_pulse}
\end{figure}
Analyzing the time evolution of the population on each
of the electronic states during an optimized pulse of 10$\,$ps pulse
duration, the molecules are found to first accumulate in the intermediate
$^1\Sigma_u^+$ and $b^3\Pi_u$ states before being dumped to the 
$X^1\Sigma_g^+$ electronic ground state. The example of an optimized
pulse with pulse duration of 8$\,$ps is presented in 
Fig.~\ref{fig:OCT_pulse}.
Inspection of the optimized pulse during a short interval of
400$\,$fs, Fig.~\ref{fig:OCT_pulse}(b), reveals that 
each peak of the pulse amplitude is correlated  to a change of
the temporal phase by $\pi$. The spectrum,
Fig.~\ref{fig:OCT_pulse}(c), displays two pronounced peaks
with maxima at $\pm$400$\,$cm$^{-1}$ with respect to the central laser 
frequency. The disappearence of the central frequency during the
optimization and appearance of two slightly detuned frequencies is
somewhat surprising since the central frequency was chosen to maximize
the effective two-photon transition  moment. There are two
possibilities to rationalize this result of the optimization: 
Either the detuning shifted by the additional $400\,$cm$^{-1}$ is
better and should be chosen for the guess pulse, or the absorption 
of two photons with different energies is more optimal than that of
two identical photons. The latter explains the observed optimal
spectrum: Analysis of the electric transition dipole moment between
the initial wave packet and the intermediate vibrational levels, 
$\langle \Psi_{in}|\Op{d_e}|v'\rangle$, reveals that it takes its 
maximal value for levels $v'$ that are detuned from the frequency
corresponding to the maximum of the effective two-photon transition
moment, Eq.~\eqref{eq:Twophoton},  
by 460$\,$cm$^{-1}$. The electric transition dipole moment between the 
intermediate and the ground state vibrational
level, $\langle v'|\Op{d_e}|v''\rangle$, attains its maximum for a
transition frequency that is smaller than the laser
frequency corresponding to the maximum of the effective two-photon
transition moment by 250$\,$cm$^{-1}$. Note that the transition
moments for absorption 
of the first photon,  $\langle \Psi_{in}|\Op{d_e}|v'\rangle$, are
about 7 times smaller than those for absorption of the second
photon, $\langle v'|\Op{d_e}|v''\rangle$. The effective two-photon
transition moment is obtained as a compromise of the two
one-photon transition moments, cf. Eq.~\eqref{eq:Twophoton} 
and Fig.~\ref{eq:Twophoton}. Allowing for two photons of different
energy in the calculation of the effective two-photon transition
moment, we still find a peak for a ground state binding energy of
2650$\,$cm$^{-1}$, which is at best 40\% higher when energies of
photons are detuned by $\pm$390$\,$cm$^{-1}$. It corresponds to
the transition frequency from the initial wave packet to the
intermediate state being 390$\,$cm$^{-1}$ larger and that between the
intermediate state and the ground electronic state being
390$\,$cm$^{-1}$ smaller than the frequency for a transition with two
identical photons. Equipped with this information, we can construct a
guess pulse that is the sum of two TL pulses with their central
frequencies separated by 780$\,$cm$^{-1}$. In this case, half of the
integrated guess pulse energy is sufficient to reach the same initial
stabilization probability, reflecting the stronger effective
two-photon transition moment. Optimization with such a guess pulse
converges faster and the final integrated pulse energy of the
optimized pulse is smaller (data not shown) but the bound for the 
stabilization efficiency, Fig.~\ref{fig:OCT_conv}, remains in place. 
The spectrum of the optimized pulse with two peaks separated by
780$\,$cm$^{-1}$ is very similar to the one shown in
Fig.~\ref{fig:OCT_pulse}, irrespective of the guess pulse central
frequency. In particular, the width of each of the peaks roughly
corresponds to the bandwidth of a TL $100\,$fs pulse. 

Optimization of the stabilization pulse reveals that the upper bound
of the stabilization efficiency, found to be 9\% for linearly chirped
pulses, can be increased up to 67\% when a TL pulse of 100$\,$fs pulse
duration is shaped and stretched to 10$\,$ps. This is significantly
more efficient than any existing proposal for short-pulse
photoassociation~\cite{KochPRA06a,KochPRA06b,KochPRA08}. 
At the same time, the 
integrated pulse energies of the optimized pulses are below 150$\,$nJ,
two to three times less than those found for the best linearly chirped
pulses or trains of TL pulses in Sec.~\ref{sec:anapulse}. The shape of
the optimized pulse is comparatively simple, characterized by a
sequence of short pulses with linear and quadratic chirps. 


\section{Summary and conclusions}
\label{sec:concl}

Based on state of the art \textit{ab initio} calculations, we have
studied the optical production of Rb$_2$ molecules in their electronic
ground state using multi-photon transitions that are driven by short
laser pulses. Our model includes not only accurate potential energy
surfaces but also spin-orbit couplings, Stark shifts and transition matrix
elements. We have employed a non-perturbative treatment of the
light-matter interaction which is crucial to capture the strong-field
effects that often accompany multi-photon transitions. 

Our proposal for the optical production of molecules using shaped
femtosecond laser pulses that drive multi-photon transitions consists,
in its first step, of
non-resonant three-photon photoassociation of atom pairs colliding
in their triplet state. 
A three-photon transition allows to access electronic states that
vary as $1/R^3$ at long range, providing comparatively large
free-to-bound transition matrix elements~\cite{KochFaraday09}. 
Strong spin-orbit interaction allows for
triplet-to-singlet conversion. The stabilization pulse, time-delayed
with respect to the photoassociation pulse, transfers the
photoassociated wave packet to the electronic ground state in a
resonant two-photon transition proceeding via the
$A^1\Sigma_u^+$-$b^3\Pi_u$ manifold. It benefits from the intricate
excited state wave packet dynamics resulting from coupled vibrational
dynamics in states with partially ion-pair character and
singlet-triplet oscillations due to the spin-orbit interaction.  

We have studied the transfer of the excited state wave packet to the
electronic ground state using transform-limited, linearly chirped and
optimally shaped laser pulses. Linearly chirped pulses
were found to perform much better, by almost an order of magnitude,
than transform-limited pulses. This is due to large Stark shifts which
drive the transition off resonance for transform limited pulses. We
have confirmed that a strong-field control scheme known for atomic
transitions, with a linear chirp compensating the phase accumulated
due to the Stark shift~\cite{TralleroPRA05,WeinachtPRL06}, can also be 
successfully employed for molecular transitions. In this case, the
chirp rate cannot be calculated analytically but needs to be
determined numerically.

Surprisingly, for optimally shaped laser pulses, the integrated
pulse energy was found to be significantly
lower than that of the best transform-limited and linearly chirped
pulses, while yielding a much better stabilization efficiency. This is
due to the fact that the transform-limited and linearly chirped laser
pulses were chosen based on an effective two-photon transition matrix
element assuming equal transition frequencies of both
photons. Optimization reveals that a two-photon transition with two
slightly different transition frequencies allows to employ two 
one-photon transitions with significantly larger transition matrix
elements. 

Overall, the stabilization efficiency is limited by somewhat
less than 70\% for transform-limited 100$\,$fs laser pulses that are
stretched to 10$\,$ps. More than 90\% transfer efficiency becomes
possible by stretching the pulse to 20$\,$ps. The target level in the
electronic ground state that is reached by these stabilization pulses
is located more than half way down the ground state potential well, with
a binding energy of about 2600$\,$cm$^{-1}$. The stabilization
efficiencies reported here have to be compared to 20\%, respectively
50\%, for stabilization with chirped pulses in the presence of strong
spin-orbit interaction to very weakly bound
levels~\cite{KochPRA06a,KochPRA06b} and to a few percent for
stabilization to the vibrational ground state using engineered
excited-state dynamics~\cite{KochPRA08}. In contrast to these earlier
studies, with the current scheme it
becomes possible to convert almost all of the
weakly bound photoassociated molecules into truly bound ground state
molecules from where a single subsequent Raman step is sufficient to
transfer these molecules into their vibronic ground state. Such a
significant improvement for comparatively short stabilization pulses
is afforded by a speed-up of the dynamics due to the partial ion-pair
character of the excited state potential curves and population
trapping at short interatomic separations due to the strong spin-orbit
interaction. It emphasizes the usefulness of multi-photon transitions,
which allow to access these electronic states,
for the photoassociation of ultracold atoms.

Multi-photon transitions moreover allow for utilizing the full
bandwidth of a 
femtosecond pulse for photoassociation and stabilization. They also
provide more flexibility with respect to the transition frequencies
that can be addressed. The basic features of femtosecond laser pulses
-- their broad bandwidth and pulse shaping capabilities -- can then be 
used to full advantage. Our results thus pave the way towards the
coherent control of photoassociation at very low temperature. 
They are not restricted to rubidium but rather are applicable to any
molecule that shows the main features of the dynamics studied here --
an ion-pair potential that is coupled to covalent states and 
strong spin-orbit interaction. 

Future work will consider optimization of the photoassociation
pulse. This represents a non-trivial control problem since the
initial state of photoassociation is the thermally populated
continuum of scattering states. While a thermal ensemble is
inherently incoherent, quantum effects are already perceptible due to
the low temperature. It will be interesting to see whether an
optimally shaped pulse can make use of the enhanced quantum purity and
presence of correlations at low temperature.

\begin{acknowledgments}
We would like to thank Ronnie Kosloff for many fruitful discussions. 
Financial support from the Polish Ministry of Science and
Higher Education through the project N N204 215539 is gratefully acknowledged.
MT is supported by the project operated within the Foundation for Polish Science MPD
Programme co-financed by the EU European Regional Development Fund.
\end{acknowledgments}

\bibliography{pa}

\end{document}